\documentclass[aps,amssymb,10pt,showpacs,showkeys,letterpaper, amssymb]{revtex4}
\usepackage{graphicx}
\usepackage{amsmath}
\usepackage{epsfig}
\usepackage{bm}

\newcommand{\be}{\begin{equation}}
\newcommand{\ee}{\end{equation}}
\newcommand{\beq}{\begin{eqnarray}}
\newcommand{\eeq}{\end{eqnarray}}

\begin{document}
\title{Crossing the phantom divide with Ricci-like holographic dark energy}

\author{ S. Lepe }
 \email{slepe@ucv.cl}
 \affiliation{\it Instituto de F\'{\i}sica, Facultad de Ciencias,
Pontificia Universidad Cat\'{o}lica de Valpara\'{\i}so, Casilla
4059, Valpara\'{\i}so, Chile}
\author{ F. Pe\~{n}a}
 \email{fcampos@ufro.cl}
\affiliation{\it Departamento de Ciencias F\'{\i}sicas, Facultad
de Ingenier\'{\i}a, Ciencias y Administraci\'{o}n, Universidad de
La Frontera, Casilla 54-D, Temuco, Chile }

\date{\today}

\begin{abstract}

In this work we study the dark energy problem by adopting an
holographic model proposed recently in the literature. In this model
it is has been postulated an energy density $\rho \sim R$, where $R$
is the Ricci scalar curvature. Under this considerations, we have
obtained a cosmological scenario which arises from considering two
non-interacting fluids along a reasonable Ansatz for the cosmic
coincidence parameter. We have adjusted the involved parameters in
the model according to the observational data and showing that the
equation of state for the dark energy exhibits a cross through the
-1 barrier. Additionally, we have found a disagreement of these
parameters in comparison with a scalar field theory approach

\end{abstract}
\maketitle
\section{Introduction}

The current observational data have suggest that the universe is
experiencing an accelerated expansion \cite{ref1, ref2, ref3}. In
order to explain this experimental evidence, many theoretical models
have been proposed in the literature. One of these models is known
as dark energy problem, which is based on an unknown fluid with a
negative pressure which drives the accelerated expansion. In this
context, many approaches have been used to describe the dark energy
such as modifications to the Einstein equations \cite{ref4}, scalar
field models and quintessence \cite{ref5}, tachyonic fields
\cite{ref6}, quintom \cite{ref7} and phantom fields \cite{ref8}. In
a recently approach \cite{ref9}, it has been proposed that the
current accelerated expansion could be explained through the vacuum
density of a colored field, responsible for a phase transition at
which the gauge SU(3)c symmetry is broken. Taking into account this
considerations, it has been studied the second order electroweak
transition, which could explain the accelerated evolution of the
universe, under certain approximations \cite{ref10}. Following this
line of reasoning, holographic approaches to explain dark energy
have also been interesting ideas to investigate \cite{ref11}. This
treatment is based on the holographic principle, which establish
that the number of degrees of freedom of a physical system should
scale with its bounding area rather than with its volume
\cite{ref12}. Many works have been developed in the literature,
using a holographic cut-off for the dark energy of the form $\rho
\sim H^{2}$ \cite{ref13} where $H$ is the Hubble parameter. However,
a recently new approach has emerged as an extension to this ideas,
in which it is proposed an holographic cut-off in the density like
$\rho \sim R$, where $R=6\left( 2H^{2}+\dot{H}+k/a^{2}\right) $ is
the Ricci scalar curvature. As a consequence, the proposed cut-off
is written as $\rho =3\left( \lambda _{1}H^{2}+\lambda
_{2}\dot{H}\right)$, where $\lambda _{1}$ and $\lambda _{2}$ are
both constants \cite{ref14}. In the literature (\cite{ref15} and
references therein), it has been considered several options for the
choose of the infrared cut-off in the holographic dark energy, such
as the Hubble parameter, the particle horizon, the future horizon as
well of any combinations of them. All these models do not consider,
at least explicitly, time derivatives terms in the Hubble parameter.
By including term of the form $\dot{H}$ it is possible to avoid the
problem of causality that is present when the future horizon is used
as a cut-off.

In our article, we will consider two non-interacting fluids: one of
them represents dark energy in accord with the proposed cut-off and
the other will play the role of usual dark matter without pressure.
The present paper is organized as follows: in Sec. II, we discuss a
Ricci-like holographic approach to the dark energy and where we have
obtained expressions for the parameters in the model in terms of
observable like the coincidence and the deceleration parameters. In
Sec. III, we shall introduce an Ansatz for the cosmic coincidence
parameter and where we show an explicit solutions for the Hubble
parameter, the cosmic scale factor and the equation of state for the
dark energy. In Sec. IV, we have obtained explicit values for
parameters involved showing that the equation of state for the dark
energy undergoes a cross through the phantom barrier. Finally, we
summarize the results in Sec. V.

\section{Holographic Ricci dark energy}

Let start our analysis by considering the non-interacting flat model
(in units $8\pi G=1$) with $\rho _{1}$ as the dark energy component
and $\rho _{2}$ the dark matter component defined by:

\begin{eqnarray}
3H^{2} &=&\rho _{1}+\rho _{2};  \label{eq1} \\
\grave{\rho}_{1} &=&3\left( \frac{1+\omega _{1}\left( z\right)
}{1+z}\right)
\rho _{1},  \label{eq2} \\
\grave{\rho}_{2} &=&3\left( \frac{1+\omega _{2}\left( z\right)
}{1+z}\right) \rho _{2},  \label{eq3}
\end{eqnarray}

where primes denote derivative with respect to the redshift
parameter defined by $1+z=\left( a/a_{0}\right) ^{-1}$ where $a$ is
the cosmic scale factor. Besides, $\omega _{1}$ and $\omega _{2}$
are parameters of the equation of state associated to $\rho _{1}$%
and\ $\rho _{2}$, respectively.

We will adopt for the dark energy component $\rho _{1}$ the
holographic Ricci approach given by:

\begin{equation}
\rho _{1}=3\left( \lambda _{1}-\lambda _{2}\left( 1+z\right) \frac{\grave{H}%
}{H}\right) H^{2},  \label{eq4}
\end{equation}

where $\lambda _{1}$ and  $\lambda _{2}$ are adjustable constants.
The additional term $\grave{H}$ included in equation (\ref{eq4})
gives a non-constant coincidence parameter, which represents a
difference with the typical $\rho _{1}\sim H^{2}$ holographic
cut-off proposed in the literature \cite{ref13} (besides, as we
stated in the introduction, this derivative term of $H$ avoids the
causality problem). We will use this fact for doing a fit of
$\lambda _{1}$ and $\lambda _{2}$ in accord to the current
observational data.

In terms of the deceleration parameter defined by:

\begin{equation}
q+1=\left( 1+z\right) \frac{\grave{H}}{H},  \label{eq5}
\end{equation}

Equation (\ref{eq4}) can be written in the form:

\begin{equation}
\rho _{1}=3\left[ \lambda _{1}-\lambda _{2}\left( q+1\right)
\right] H^{2}. \label{eq6}
\end{equation}

By combining of equations (\ref{eq6}) and (\ref{eq1}), the energy
density $\rho _{2}$ becomes:

\begin{equation}
\rho _{2}=3\left[ \left( 1-\lambda _{1}\right) +\lambda _{2}\left(
q+1\right) \right] H^{2},  \label{eq7}
\end{equation}

From equations (\ref{eq6}) and (\ref{eq7}) it is possible to
establish the following constraint for the deceleration parameter
$q\left( z\right)$:

\begin{equation}
\frac{\lambda _{1}}{\lambda _{2}}-\frac{1}{\lambda _{2}}<q\left( z\right) +1<%
\frac{\lambda _{1}}{\lambda _{2}}, \label{eq8}
\end{equation}

As a consequence that both $\rho _{1}$ and $\rho _{2}$ satisfy the
weak energy condition. Besides, the constraint indicates that
$q\left( z\right)$ have a strong dependence on the allowed values of
$\lambda _{1}$ and $\lambda _{2}$, considered either by theoretical
models or by observational data.

- The coincidence parameter.

The coincidence parameter is defined by the ratio:

\begin{equation}
r=\frac{\rho _{2}}{\rho _{1}}=\frac{\left( 1-\lambda _{1}\right)
/\lambda _{2}+\left( q+1\right) }{\lambda _{1}/\lambda _{2}-\left(
q+1\right) }. \label{eq9}
\end{equation}

By using the previous definition, it is possible to express the deceleration parameter
$q(z)$ in terms of coincidence parameter $r$ by:

\begin{equation}
q\left( z\right) +1=\frac{1}{\lambda _{2}}\left[ \lambda
_{1}-\left( 1+r\left( z\right) \right) ^{-1}\right],
\label{eq10}
\end{equation}

The current observational data have shown that $q\left( 0\right)<0$.
This condition imposes a constrain over the possible values of the
adjustable constants of the model, given by:

\begin{equation}
\lambda _{1}-\lambda _{2}<\left[ 1+r\left( 0\right) \right] ^{-1},
\label{eq11}
\end{equation}

By combining equations (\ref{eq10}) and (\ref{eq11}), this constrain
takes the following form:

\begin{equation}
\left( \lambda _{1}-\lambda _{2}\right) -\left[ 1+r\left( 0\right)
\right] ^{-1}=\lambda _{2}q\left( 0\right) <0\Longrightarrow
\lambda _{2}>0, \label{eq12}
\end{equation}

and from this relationship, together with equation (\ref{eq10}), we
have also obtained $\lambda _{1}>0$. Then, by combining equations
(\ref{eq10}) and (\ref{eq12}) it is possible to obtain the following
expression for the adjustable parameters of the present model:

\begin{equation}
\lambda _{1}=\frac{1}{1+r\left( \infty \right) }+\frac{1+q\left( \infty
\right) }{q\left( \infty \right) -q\left( 0\right) }\left[ \frac{1}{%
1+r\left( 0\right) }-\frac{1}{1+r\left( \infty \right) }\right] ,
\label{eq13}
\end{equation}

and

\begin{equation}
\lambda _{2}=\frac{1}{q\left( \infty \right) -q\left( 0\right) }\left[ \frac{%
1}{1+r\left( 0\right) }-\frac{1}{1+r\left( \infty \right) }\right]
, \label{eq14}
\end{equation}

With the previous expressions and by using the current observational
data, it is possible to fix the values $\lambda _{1}$ and $\lambda
_{2}$ for our model. If we consider, at early times, the cosmic
evolution driven by dark matter like dust, the value $q\left( \infty
\right) $ is equal to $1/2$ .

On other hand, the acceleration of the expansion is given by:

\begin{equation}
\frac{\ddot{a}}{a}\left( z\right) =-q\left( z\right) H^{2}\left(
z\right) , \label{eq15}
\end{equation}

By using the expression for the deceleration parameter given in
equation (\ref{eq10}), it can be written in the form:

\begin{equation}
\frac{\ddot{a}}{a}\left( z\right) =\frac{1}{\lambda _{2}}\left[ \frac{1}{%
1+r\left( z\right) }-\left( \lambda _{1}-\lambda _{2}\right)
\right] H^{2}\left( z\right) ,  \label{eq16}
\end{equation}

and expression (\ref{eq16}) reveals that only a positive
acceleration is consistent with the constrain given in equation
(\ref{eq12}), i. e., there is no transition from a decelerated
regime to an accelerated one.

-The equation of state.

In what follow, we will analyze the equation of states in our model.
Let consider the definitions for the dark energy component $\rho
_{1}$ and the dark matter $\rho _{2}$ given in equations (\ref{eq2})
and (\ref{eq3}) together with the definition of the coincidence
parameter $r$ given by equation (\ref{eq9}). By combining these
equations, we obtain the following expression:

\begin{equation}
\omega _{1}-\omega _{2}=-\frac{1}{3}\left( 1+z\right)
\frac{\grave{r}}{r}, \label{eq17}
\end{equation}

By using equation (\ref{eq1}), (\ref{eq9}) and (\ref{eq5}), it is
straightforward to show that:

\begin{eqnarray}
1+\omega _{1}\left( z\right) &=&\frac{2}{3}\left( q\left( z\right) +1\right)
-\frac{1}{3}\left( 1+z\right) \frac{\grave{r}\left( z\right) }{1+r\left(
z\right) },  \label{eq18} \\
1+\omega _{2}\left( z\right) &=&\frac{2}{3}\left( q\left( z\right)
+1\right) +\frac{1}{3}\left( 1+z\right) \frac{\grave{r}\left(
z\right) }{r\left( z\right) \left[ 1+r\left( z\right) \right] }.
\label{eq19}
\end{eqnarray}

Considering $\omega _{2}\left( z\right) =0$ for the dark matter, we
obtain:

\begin{equation}
\grave{r}\left( 0\right) =\left[ 1-2q\left( 0\right) \right]
r\left( 0\right) \left[ 1+r\left( 0\right) \right] ,  \label{eq20}
\end{equation}

so that $\grave{r}\left( 0\right) $ it is determined from the
observational data for $q\left( 0\right) $ and $r\left( 0\right) $.

From equations (\ref{eq18}) and (\ref{eq20}) it is possible to
obtain an expression for $\omega _{1}\left( 0\right)$ in the form:

\begin{equation}
\omega _{1}\left( 0\right) =-\frac{1}{3}\frac{\grave{r}\left( 0\right) }{%
r\left( 0\right) }=-\frac{1}{3}\left[ 1-2q\left( 0\right) \right]
\left[ 1+r\left( 0\right) \right] .  \label{eq21}
\end{equation}

The set of equation of states given by (\ref{eq18}) and
(\ref{eq19}), together with the definition of the deceleration
parameter given in equation (\ref{eq5}), show explicitly the
inhomogeneous character of our model for the dark energy problem
\cite{ref16}.

In the next section, after choosing a reasonable Ansatz for the
coincidence parameter, we will give an explicit solutions for
$H\left( z\right)$ and the cosmic scale factor and we will discuss
the cosmology in this model.

\section{An Ansatz for the coincidence parameter}

Let consider the following Ansatz for the coincidence parameter

\begin{equation}
r\left( z\right) =r_{0}+\epsilon _{0}z\left( 1+z\right) ^{-1},
\label{eq22}
\end{equation}

where $r\left( 0\right) =r_{0}$ , $\grave{r}\left( 0\right)
=\epsilon _{0}$ and $r\left( \infty \right) =r_{0}+\epsilon _{0}$

This type of parametrization has been considered in previous works
concerning to an interacting scheme between dust and a holographic
dark energy density described by $\rho _{1}\sim H^{2}$)
\cite{ref17}.

Starting from the definitions given in equations (\ref{eq1}) and
(\ref{eq4}) and by using the proposed Ansatz, it is possible to
obtain the following solution for the Hubble parameter

\begin{eqnarray}
H\left( z\right) &=&H\left( 0\right) \left( 1+z\right) ^{\lambda
_{1}/\lambda _{2}}\left[ \left( \frac{1+r_{0}}{\epsilon _{0}}\right) \frac{%
1+z_{s}}{z-z_{s}}\right] ^{1/\lambda _{2}\left( 1+r_{0}+\epsilon _{0}\right)
};  \nonumber \\
1+z_{s} &=&\epsilon _{0}\left( 1+r_{0}+\epsilon _{0}\right) ^{-1},
\label{eq23}
\end{eqnarray}

where it is straightforward to verify the following limit:

\begin{equation}
H\left( z\rightarrow \infty \right) \rightarrow \left( 1+z\right)
^{\left[ \lambda _{1}-1/\left( 1+r_{0}+\epsilon _{0}\right)
\right] /\lambda _{2}}, \label{eq24}
\end{equation}

so that,

\begin{eqnarray}
H\left( z\rightarrow \infty \right) &\rightarrow &\infty \Longleftrightarrow
\lambda _{1}>\left( 1+r_{0}+\epsilon _{0}\right) ^{-1},  \label{eq25} \\
&\Longrightarrow &\left( 1+r_{0}+\epsilon _{0}\right)
^{-1}<\lambda _{1}<\lambda _{2}+\left( 1+r_{0}\right) ^{-1},
\label{eq26}
\end{eqnarray}

where the inequality given in equation (\ref{eq11}) has been used to
obtain the expression (\ref{eq26}).

From equations (\ref{eq22}) and (\ref{eq23}), it is possible to
obtain the following expression for the Hubble parameter:

\begin{equation}
\frac{\grave{H}}{H}\left( z\right) =\frac{1}{\lambda _{2}}\left[ \frac{%
\lambda _{1}}{1+z}-\frac{1}{\epsilon _{0}}\left( \frac{1+z_{s}}{z-z_{s}}%
\right) \right] ,  \label{eq27}
\end{equation}

By using the equations (\ref{eq23}) and (\ref{eq27}), the
acceleration can be written in the form:

\begin{equation}
\frac{\ddot{a}}{a}\left( z\right) =H^{2}\left( z\right) \left[ 1-\frac{1}{%
\lambda _{2}}\left( \lambda _{1}-\frac{1}{\epsilon
_{0}}\frac{\left( 1+z_{s}\right) \left( 1+z\right)
}{z-z_{s}}\right) \right] ,  \label{eq28}
\end{equation}

Finally, by using equations (\ref{eq18}), (\ref{eq23}) and
(\ref{eq27}) it is possible to write the equation of state
corresponding to $\omega _{1}$ in the form:

\begin{equation}
1+\omega _{1}\left( z\right) =\frac{2}{3\lambda _{2}}\left[ \lambda _{1}-%
\frac{1}{\epsilon _{0}}\frac{\left( 1+z_{s}\right) \left( 1+z\right) }{%
z-z_{s}}\right] -\frac{1}{3}\frac{\epsilon _{0}\left( 1+z_{s}\right) }{%
\left( 1+r_{0}\right) \left( 1+z_{s}\right) +\epsilon _{0}z},
\label{eq29}
\end{equation}

and

\begin{equation}
1+\omega _{1}\left( z\rightarrow \infty \right) \rightarrow \frac{2}{%
3\lambda _{2}}\left[ \lambda _{1}-\left( 1+r_{0}+\epsilon _{0}\right) ^{-1}%
\right] -\frac{1}{3}\left( \frac{\epsilon _{0}}{1+r_{0}+\epsilon _{0}}%
\right) \frac{1}{z}.  \label{eq30}
\end{equation}

If we analyze the expression obtained for the Hubble parameter, it
is possible to notice that exist a singularity at the value $
z=z_{s}=\epsilon _{0}\left( 1+r_{0}+\epsilon _{0}\right) ^{-1}-1$.
This is a type III singularity: at a finite value of the scale
factor both energy density and pressure diverge \cite{ref18}.
However, this singularity does not occur in the future evolution
given that, before this point is achieved, the coincidence parameter
given in equation (\ref{eq22}) has vanished ( $r\left(
\overline{z}\right) =0$ for $\overline{z}=\epsilon _{0}\left(
r_{0}+\epsilon _{0}\right) ^{-1}-1>z_{s}$). Therefore, the evolution
cross the phantom divide but no experience a future singularity.

A final calculations, just for completeness, gives the solutions for
the scale factor when $z\rightarrow\infty$ and $z\rightarrow z_{s}$.
The expressions are, respectively

\begin{equation}
a\left( t\right) =a\left( 0\right) \left[ \alpha ^{\beta }\left( \lambda
_{1}/\lambda _{2}-\beta \right) H\left( 0\right) \left( t-t_{0}\right) +1%
\right] ^{1/\left( \lambda _{1}/\lambda _{2}-\beta \right)
}\Longrightarrow \ddot{a}\left( t\right) >0  \label{eq31}
\end{equation}

where we have used equation (\ref{eq11}) and the definitions given
by:

\begin{equation}
\alpha =\left( \frac{1+r_{0}}{\epsilon _{0}}\right) \left(
1+z_{s}\right) \text{ \ \ },\text{ \ \ }\beta =\frac{1}{\lambda
_{2}\epsilon _{0}}\left( 1+z_{s}\right) .  \label{eq32}
\end{equation}

and

\begin{equation}
a\left( t\right) =a\left( t_{s}\right) -a\left( 0\right) \left[
\left( \beta +1\right) C\right] ^{1/\left( \beta +1\right) }\left(
t_{0}+t_{s}-t\right) ^{1/\left( \beta +1\right) }  \label{eq33}
\end{equation}%

where $C$ is a constant and $t_{s}\sim \left[ a\left( t_{s}\right)
/a\left( 0\right) -1\right] ^{\beta +1}.$

\section{The observational data}

In order to fit the adjustable parameter of our model, $\lambda
_{1}$ and $\lambda _{2}$, we have used the following observational
data: $r_{0}\approx 0.37$ from Refs. \cite{ref1, ref2} and
$q_{0}\approx -0.7$ from Ref. \cite{ref3}. By replacing these values
in equation (\ref{eq20}) we get the value $\epsilon _{0}\approx
1.22$. Under these consideration, the adjustable parameters take the
values $\lambda _{1}\approx 0.8$ and $\lambda _{2}\approx 0.3$.
Additionally, we have obtained the following limits for the equation
of state $\omega _{1}\left( z\rightarrow \infty \right) \approx
-0.08\rightarrow $ $\omega _{1}\left( 0\right) \approx -1.096$. In
this sense, we can pointed out as a prediction of our model, that
the equation of state of the dark energy cross the phantom divide.
This type of behavior was obtained before, by using other approaches
such as quintom cosmology \cite{ref7}, nevertheless, as a difference
and an advantage in comparison with the quinton approach, our model
does not required a phantom field as an input.

Finally, it is possible to compare our results with those ones
reported in Ref. \cite{ref19}. In this work, authors described the
same problem, but considering a scalar field theory based on quantum
vacuum fluctuations as a physical constituent of dark energy, in
order to visualize the constants $\lambda _{1}$ and $\lambda _{2}$
under a dynamical scope. They claim that the origin of
$\rho =3\left( \lambda _{1}H^{2}+\lambda _{2}\dot{H}%
\right) $ in the framework of the holographic principle is in a
sense kinematical, i.e., it lacks any dynamical support. The
constants $\lambda _{1}$ and $\lambda _{2}$ are, in principle,
arbitraries and in this sense, without dynamics. Besides, they have
obtained the following expression for the adjustable parameter
$\lambda _{1}=N/8\pi -2$ and $\lambda _{2}=N/16\pi -2$ where $N$ is
the difference between the bosonic and fermionic fundamental modes
present in the theory. They have calculated that the value $N\approx
100$ is a quite realistic result if it is taken the limit $\omega
_{1}\sim -1$. However, according to our fitted values for $\lambda
_{1}$ and $\lambda _{2}$, it is possible to obtain the value $
N\approx 25$. Thus, our fitted set of parameter $\left\{ \lambda
_{1},\lambda _{2}\right\} $ is in clear disagreement with those ones
which comes from the scalar field theory aforementioned.

\section{Final remarks}

In this work we have discussed a holographic model for the dark
energy component inspired in $\rho _{1}\sim R$  where $R=6\left(
2H^{2}+\dot{H} +k/a^{2}\right)$ is the Ricci scalar curvature. We
have obtained from it an early quintessence behavior for the dark
energy and a late phantom evolution, that is, a quintom-like scheme.
A reasonable Ansatz for the coincidence parameter was used and in
terms of it we found explicit values for the parameters which
characterizes the model. Although the predictions of the
$\Lambda$CDM model have a remarkable consistency with the current
observational data, the dynamical model that we have used, it is
mildly favored by the observations because it is possible to explain
the barrier -1 crossing.

Finally, we did not found a good agreement with a scalar field
theory approach about the values of these parameters.

\begin{acknowledgments}

One of us (SL) acknowledge the hospitality of the Physics Department
of Universidad de La Frontera where part of this work was made. This
work was supported from DIUFRO DI10-0009, Direcci\'{o}n de
Investigaci\'{o}n y Desarrollo, Universidad de La Frontera (FP).

\end{acknowledgments}

\end{document}